**Growth of bulk single-crystal MnP helimagnet and its structural and NMR characterization**

Nikolai D. Zhigadlo[a,*], Nicolo Barbero[b], Toni Shiroka[b,c]

[a] *Department of Chemistry and Biochemistry, University of Bern, Freiestrasse 3, 3012 Bern, Switzerland*

[b] *Laboratory for Solid State Physics, ETH Zurich, Otto-Stern-Weg 1, 8093 Zurich, Switzerland*

[c] *Paul Scherrer Institut, CH-5232 Villigen PSI, Switzerland*

(Dated: May 26, 2017)

**Abstract**

Bulk single crystals of manganese phosphide (MnP) were grown from melt at 1 GPa and 1200 °C by using a cubic-anvil, high-pressure, and high-temperature technique. The obtained black colored crystals exhibit a plate-like morphology, with flat surfaces and maximum dimensions up to ~ 4 × 2 × 0.5 mm$^3$. The orthorhombic crystal structure was confirmed by X-ray diffraction [*Pnma*, 62, $Z = 4$, $a = 5.2510(4)$ Å, $b = 3.1670(3)$ Å, $c = 5.90098 (4)$ Å and V = 98.279(14) Å$^3$]. Temperature-dependent magnetization measurements reveal the occurrence of two successive transitions: a paramagnetic to ferromagnetic transition at $T_c = 290.5$ K and the development of a double helimagnetic order at $T_s = 44.5$ K. Zero-field $^{31}$P NMR measurements in the FM and in the screw-spin AFM state show prominent features, which are compared with previous experimental data and theoretical calculations. The relatively large crystals obtained here open up new possibilities for further explorations of this interesting material.

*Corresponding author.

*E-mail address*: nikolai.zhigadlo@dcb.unibe.ch or nzhigadlo@gmail.com  (N. D. Zhigadlo)

## 1. Introduction

The MX binary compounds (M: transition metal, X: pnictogen) with an MnP-type structure have constantly attracted researchers' attention due to a variety of magnetic orderings and to possible applications in areas such as spintronics, magnetostriction, and magnetic refrigeration. For instance, the first-order ferromagnetic transition in MnP, occuring near room temperature, gives rise to a pronounced magneto-caloric effect (MCE), ranking MnP and its substituted variants among the most promising MCE materials [1]. However, the most striking property of MnP and the related compounds CrAs and FeP [2-7] is arguably their helimagnetism. Despite their apparently simple chemical formulae, due to an intrinsic structural complexity [8], these materials exhibit unusual helical magnetic structures, which are still a matter of debate [9,10]. The archetypal MnP compound has four formulas units per unit cell [Fig. 1] and consists of $Mn^{3+}$ manganese ions occupying equivalent crystal sites, surrounded by distorted $P_6$ octahedra, to form $MnP_6$ units.

Depending on temperature, magnetic field, and pressure, MnP exhibits a rich magnetic phase diagram, as discussed extensively in literature [11-17]. At ambient pressure and without an applied field, MnP shows two successive magnetic transitions as the temperature is decreased. A first para- to ferromagnetic (FM) transition occurs at $T_c \approx 291$ K which, below $T_s \approx 47$ K, transforms into a double helical antiferromagnetic (AFM) order with propagation vector $\boldsymbol{q} = (0,0,0.117)$ [13]. MnP is a primary example of a magnetic system with a Lifshitz point, i.e., a system where the boundaries of paramagnetic, ferromagnetic, and helicoidal (fan) phases meet at a triple point [18]. This fact initially triggered a considerable theoretical and experimental interest, which has persisted over the years. Very recently it was found that, under high-pressure conditions, both CrAs and MnP become superconductors [19-22], whose nature resembles that of unconventional superconductors like cuprates, heavy fermions, and iron pnictides [23,24]. In all these systems superconductivity emerges in the vicinity of a magnetic quantum critical point, where magnetism is suppressed. Whether the newly discovered MnP superconductivity is indeed of unconventional type or not is still being debated [25]. To understand the superconductivity paring mechanism, it is crucial to elucidate the character of the magnetic phase preceding it. For the MnP case the situation is significantly more complicated, since various magnetic phases compete with each other until the superconductivity sets in at ~8 GPa.

Recently, the evolution of the MnP magnetic state under high pressure was investigated via muon-spin spectroscopy [26]. These studies found that pressure suppresses rather quickly the helical state and reduces the $T_c$ of the FM transition. Furthermore an AFM-like phase seems to emerge above ~1.5 GPa, successively completely disappearing above 8 GPa. All these transitions are characterized by coexisting FM and AFM phases, observed both as a function of temperature and pressure. The detailed structure of the new AFM order, which precedes the appearance of superconductivity, is still unclear and, hence, deserves further study.

In order to understand the nature of the magnetic phase bordering with superconductivity, the availability of bulk high-quality single crystals of MnP is essential. Single-crystal studies are obviously preferred to studies on polycrystalline materials, since the former bypass the complications arising from the MnP anisotropy. The measurement of anisotropic properties is interesting on its own, as it can serve as an indicator of crystal symmetry and its changes. In previous works, single MnP crystals were produced either by using a temperature-gradient furnace technique or from Sn or Se fluxes [22]. Both these methods typically lead to needle-like crystals, unsuitable for many interesting physical properties measurements. Here, we propose an alternative way, based on high-pressure, high-temperature methods, to achieve bulk single crystals of relatively large sizes, which open up new possibilities for further exploring these interesting materials. Besides the details of crystal growth, below we report also about their structural, magnetic, and NMR characterization.

2. **Experimental details**

For the growth of MnP single crystals, we utilized the cubic-anvil high-pressure and high-temperature (HPHT) technique. The latter was developed earlier on for growing superconducting intermetallic crystals [27,28], cuprate oxides [29], $Ln$Fe$Pn$O ($Ln$1111, $Ln$: lanthanide, $Pn$: pnictogen) oxypnictides [30-33], and numerous other compounds [34]. Our apparatus is equipped with a hydraulic-oil system comprising a semi-cylindrical multi-anvil module (Rockland Research Corp.) [Fig. 2(a)]. The latter consists of eight large outer steel dies squeezing six small inner tungsten carbide anvils (with edge length 22 mm), thus applying an amplified pressure on an experimental cubic cell of ~27 × 27 × 27 mm$^3$ in size [Fig. 2(b)]. The assembly has a single-stage graphite tubular heater (12 mm in inner diameter and 22.3 mm in length) placed in the bore of a pyrophyllite cube. The heater is filled with a cylindrical boron nitride (BN) crucible in the center and with pyrophyllite tablets on top and bottom [Fig. 2(c)]. A current passing through the graphite heater, supplied from outside through the steel parts and the two tungsten carbide anvils, is used for heating. The temperature is calibrated in advance and is related to the power dissipated in the pressure cell. To avoid overheating the tungsten carbide anvils, a water cooling system is installed.

A mixture of stoichiometrically equal amounts of manganese powder (purity 99.99%) and red phosphorus powder (purity 99.999%) was thoroughly grounded and pressed into a pellet of approximately 8 mm in diameter and 8 mm in length. The pellet was then placed in a BN container surrounded by a graphite-sleeve resistance heater and inserted into a pyrophyllite cube. All the work related to the sample preparation and the packing of the high-pressure cell-assembly was performed in a glove box with a protective argon atmosphere. After completing the crystal growth process, the MnP crystals were mechanically extracted from the solidified lump.

Single crystal X-ray diffraction was performed on an *Oxford Diffraction SuperNova* area-detector diffractometer [35] using mirror optics, monochromatic Mo $K_\alpha$ radiation ($\lambda = 0.71073$ Å), and an Al filter [36]. The unit cell parameters and the orientation matrix for the data collection were obtained from a least-squares refinement, using reflection angles in the range $5.1 < \theta < 27.8°$. A total of 1009 frames were collected using $\omega$ scans, with 1.0+1.0 seconds of exposure time, a rotation angle of 1.0° per frame, and a crystal-detector distance of 65.0 mm at $T = 298(2)$ K. The data reduction was performed using the *CrysAlisPro* [35] program. The intensities were corrected for Lorentz- and polarization effects, and an absorption correction based on the multi-scan method using SCALE3 ABSPACK in *CrysAlisPro* [35] was applied. The elemental analysis of the grown products was performed by means of Energy Dispersive X-ray spectroscopy (EDX, Hitachi S-3000 N). The temperature-dependent magnetization measurements were carried out using a Magnetic Property Measurement System (MPMS-XL, Quantum Design) equipped with a reciprocating-sample option.

Microscopically the newly synthesized compound was studied via zero-field (ZF) nuclear magnetic resonance (NMR) in both the ferromagnetic (FM) and antiferromagnetic (AFM) phases. The compound Mn(III)P has two single P and Mn sites, hosting respectively $^{31}$P nuclei with spin $I = 1/2$ and quadrupole $^{55}$Mn nuclei with spin $I = 5/2$. Since the former exhibit a much stronger

NMR signal (due to a large gyromagnetic ratio), we used $^{31}$P as the probe of choice for investigating the magnetism of MnP. The NMR measurements included line-shapes, spin-lattice relaxation times ($T_1$), and spin-spin relaxation times ($T_2$). The spectra were obtained via Fourier transformations of the spin echoes, derived from a standard Hahn $\pi/2$-$\pi$ sequence (for the line-shapes and $T_2$ measurements) and from an inversion recovery sequence ($\pi$-$\pi/2$-$\pi$) for the $T_1$ measurements. Depending on temperature, pulse widths of 3-4 µs and echo delays of 30 µs with repetition times from 2 to 100 ms (increasing upon cooling) were used. To cover the rather broad frequency range of the NMR lines (several MHz), the line-shapes were measured via frequency sweeps using 30-kHz steps.

## 3. Results and discussion

The HPHT method is very effective in synthesizing new inorganic materials. However, due to the limited availability of high-pressure phase diagrams, it remains mostly an empirical technique based on rational trial and error. For the high-pressure synthesis of MnP, the cubic anvil cell illustrated in Fig. 2(b,c) was brought to 1 GPa at room temperature and the optimum growth conditions were tuned by varying the heating temperature, the reaction time, and the cooling rate. After this preliminary optimization, we used a synthesis temperature of about 1200 ºC, which was found to be optimal for growing sizable MnP single crystals. In a typical growth process the BN crucible was heated up to ~1200 ºC in 3 h and maintained there for 2 h. Then the melt was slowly cooled to 950 ºC over 15 h and finally it reached room temperature in another 2 h. The product inside the crucible was a fully melted cylindrically-shaped solidified lump. Such an as-grown piece was mechanically split along its radial direction to generate specimens as those shown in Fig. 3(a-d). The extracted black coloured MnP crystals were found to exhibit irregular plate-like shapes with flat and slightly cryptomorphic surfaces reaching a maximum dimensions of ~4 × 2 × 0.5 mm$^3$ [Fig. 3(a-d)]. Basically the crystal size was limited by the crucible; in fact the circular cross section shown in Fig. 3(a,b) reflects the inner crucible walls. These crystals were free from visible inclusions and cracks, as shown in the magnified SEM image [Fig. 3(d)] of the crystal presented in Fig. 3(c). The EDX analysis of the whole solidified matrix shows a homogeneous distribution of Mn and P atoms. The ratio of Mn-to-P atoms was determined by averaging several measurements [Fig. 3(e)] and was found to be 1:1, consistent with the stoichiometric composition (within instrumental error). Consequently, the desired MnP phase represents the whole volume of the crucible. These results suggest that MnP is a compound which melts congruently under high-pressure conditions. Temperature-dependent magnetization measurements revealed the occurrence of two successive transitions with decreasing temperature (see Fig. 4): from paramagnetic to ferromagnetic state at $T_c$ = 291.5 and a second one to a double helimagnetic order at $T_s$ = 44.5 K, in good agreement with previous reports [26]. Further examinations of the plate-like shaped crystals revealed that they belong to the *Pnma* space group.

Before entering into the details of high-pressure growth we note that, when MnP crystals are grown by the temperature-gradient furnace technique or from Sn or Se fluxes [22], they often exhibit needle-shaped crystals and require long times (even several days) to reach millimeter-sized dimensions. Usually, the Sn and Se fluxes are filtered by centrifuging them before the ampoule is cooled to room temperature and then washed with diluted acid. These procedures, nevertheless, do not guarantee that some residual amount of flux is still incorporated into the crystals. The high-pressure synthesis reported here eliminates the above drawbacks completely.

Currently it is not easy to put forward a precise mechanism for the high-pressure growth of MnP crystals. However, from the basic principles of HPHT crystal growth and by taking into account the current and previously published results, we can propose some possible explanations. In our growth attempts we used a stoichiometric ratio for the reagents, of which the liquid phosphorus most probably works as a flux promoting the solution of the Mn constituent. We used red phosphorus as a starting reagent, known to be stable below 500 °C at 1 GPa, whereas the powdered black phosphorus, a polymorphic transformation of red phosphorus, appeared between 550 and 1000 °C [37]. At 1 GPa and 1050 °C the liquid phase of black phosphorus (with orthorhombic structure) appears, which consists of tetrahedral $P_4$ molecules [37]. Interestingly, a further moderate compression or heating leads to a liquid-liquid phase transformation, where tetrahedral $P_4$ molecules adopt a polymeric form, in which atoms are connected by anisotropic bonds leading to a network configuration [38,39]. Our MnP crystal-growth attempts were conducted precisely in the conditions where both P liquids co-exist. Furthermore, it is known that the ambient-pressure melting point of MnP is about 1140 °C [1]. Thus, the combined 1 GPa - 1200 °C condition seems very favorable for growing bulk MnP single crystals. To further clarify the mechanism of MnP growth one needs to know the detailed pressure-temperature phase diagram of the Mn-P system, which currently is not available.

Let us now focus on the structural properties of the grown crystals. The first MnP crystal structure was reported in the 60's by Rundqvist [11]. To check the structure of our crystals we collected diffraction data on several MnP pieces, originating from different growth batches. For each case, a useful single crystal could be found and the refined structural model was essentially equivalent to the reported one [11].

A full X-ray refinement was performed on a crystal with dimensions $0.2 \times 0.17 \times 0.13$ mm$^3$, using 1334 reflections (of which, 136 unique) in the $k$-space region $-6 \leq h \leq 6$, $-4 \leq k \leq 4$, $-7 \leq l \leq 7$, by a full-matrix least-squares minimization of $F^2$. The weighting scheme was based on counting statistics and included a factor to down-weight the most intense reflections. The structure refinement was performed using the SHELXL-2014 program [40], in the orthorhombic space group *Pnma*. Figure 1 shows the 3D view of the crystal structure of MnP (*Pnma*, 62) with unit cell dimensions $a = 5.2510$ Å, $b = 3.1670(3)$ Å and $c = 5.9098(4)$ Å at room temperature. The Mn and the P element occupy the 4$c$ ($x$, 1/4, $z$) crystallographic positions with $x_{Mn} = 0.0047(1)$, $z_{Mn} = 0.1966(1)$ and $x_P = 0.1880(1)$, $z_P = 0.5686(1)$. The crystallographic parameters are summarized in Table 1. The refined atomic positions, bond lengths, and angles are presented in Table 2 and Table 3. The full thermal displacement parameters $U^{ij}$ for all atoms are reproduced in Table 4. The $R_1$ refinement factor was 1.86% with w$R_2$ = 4.41% including all data. The chemical formula of MnP was confirmed from the fully occupied positions for all sites. No additional phases (impurities, twins, or intergrowth crystals) were detected in the reconstructed reciprocal space sections of the MnP single crystals.

The microscopic nature of the MnP magnetism was explored by $^{31}$P NMR measurements in the FM and AFM regions. Data from Nagai *et al*. [41] were used as an experimental reference, while theoretical calculations of Winter [42] served as a theoretical framework. Starting from the high-temperature paramagnetic (PM) phase, the already remarkable Knight shift $K_s$ of the $^{31}$P nucleus at 375 K (-11% – see Ref. [43],) seems to decrease even more dramatically as $T_c$ is approached from above. Therefore, an NMR study of the paramagnetic phase occurring for $T > T_c$ is possible only at high temperatures (well above 500 K). The Clogston-Jaccarino plot [i.e., the $K_s(^{31}$P) shift vs. $\chi$ susceptibility plot – not shown] exhibits a linear trend with a slope $\alpha$ = -6.70 (emu/mol)$^{-1}$, which implies an isotropic hyperfine field $H_{hf}$ = -3.7 T in the PM phase, in good agreement with the data reported in Ref. [41]. Regarding the magnetically-ordered phases, the compound Mn(III)P exhibits a *Pnma* space group with $Z$ = 4 and only a single (4$c$) site for the

spin-1/2 $^{31}$P nucleus. Nevertheless, in both the FM and AFM phases, the observed zero-field $^{31}$P NMR lineshape consists of two peaks (instead of the expected single peak), respectively labelled with L (low-frequency) and H (high-frequency) – see Fig 5. The line splitting reflects essentially the different local magnetic fields at the $^{31}$P sites. These fields arise, respectively, from two distinct contributions: those from domains, where the Mn electronic moments point along the *c* axis (H) and those from domain walls (Bloch walls - L), where the Mn electronic moments point along the *b* axis [42]. The temperature dependence of both the L- and H-frequencies for selected cases is shown in Fig. 6. Both frequencies are compatible with a hyperfine internal field of -4.4 T (see below) and decrease smoothly as the temperature increases towards $T_c$. At low temperatures, the magnetization curves follow a $\Delta M/M_0 = AT^\beta$ power-law, with $\beta = 1.47(5)$ – very close to the 3/2-value expected for spin-wave excitations. Interestingly, unlike the macroscopic magnetization which exhibits a major jump across the heli- to ferromagnetic phase transition at $T_s$ (see Fig. 4), the NMR frequencies show only minor drops at $T_s$, more prominent for the L-line. Hence, the transition between the two magnetically-ordered phases, does not imply significant microscopic changes to the internal field, despite the dramatic macroscopic outcome. This result is in stark contrast with the PM-to-FM transition, where both micro- and macroscopic probes sense major magnetization changes.

The existence of two distinct contributions to the H and L lines has been confirmed through similar NMR experiments on single crystals [41]. In particular, the intensity *I* of the $^{31}$P line has been measured in different applied magnetic fields $H_A$ where, instead of following the expected trend within a magnetic domain, i.e., $I \propto 1/(H+H_A)^2$, it was shown that the intensity decreases more rapidly and with differing rates for the L and H lines. The slower intensity decay for the L lines is due to the resonance enhancement within the Bloch walls, a fact demonstrated theoretically in Refs. [42] and [44]. From our line splitting, we estimate an $H_{hf} = -4.4$ T and, from $K_s$ calculations, we deduce a $f_{ref} = 75.8326$ MHz. The high-frequency H lines become slightly broader upon cooling, especially below $T_s$, since the magnetic domains experience an increased hyperfine and anisotropic field when going from the FM to the AFM phase (see Fig. 5). On the other hand, the L lines become narrower, since the anisotropy energy limits the thickness of the Bloch walls and, hence, the reorientation of the magnetic moments occurs less gradually (i.e., within fewer atomic planes).

The $T_1$ measurements confirm that the L lines correspond indeed to contributions from the domain (Bloch) walls and that the H lines correspond to contributions from the domains. In fact, as predicted in Ref. [42], spin-lattice relaxation processes occurring in the domain walls are faster than those in the domains. By measuring the $T_1$ vs. $T$ dependence for the H line, we observe a constant $T_1T = 0.24 \pm 0.01$ sK away from the peak at $T_s$ (see Fig. 7). On the other hand, we were unable to measure the $T_1$ values related to the L line. From this we infer that L lines have a faster dynamics than the minimum delay we could use (~10 μs). Neutron diffraction measurements [16] confirm that the gradual reorientation of the magnetic moments occurs in the domain walls, which are only of the 180°-type.

The temperature dependences of the spin-lattice $T_1$ and spin-spin $T_2$ (the last not shown) follow the typical trends observed in solids, with the first decreasing linearly and the second increasing moderately upon heating, reflecting correlation times inversely proportional to temperature. Our $T_1$ values are reported in Fig. 7 and show a very prominent peak across $T_s$; $T_2$ values instead were found to vary less, ranging from 50 μs at 5 K to 70 μs at 100 K. It is worth mentioning that the stretching parameter β (used to model the exponential recovery of the longitudinal magnetization and typically around 1) has values of about 0.5 (see inset in Fig. 7). This rather small value could reflect the significant magnetic anisotropy and the fact that (especially in no applied field) one can still have multiple magnetic domains within a single crystal. The integration over all the possible orientations produces then a stretched-exponential relaxation, which justifies the small experimental β values we find.

## 4. Conclusion

A new method to grow single crystals of the MnP helimagnet is presented. By using a wedge-type, cubic-anvil, high-pressure apparatus, bulk single crystals of MnP were grown from stoichiometric melts at 1200 °C under a pressure of 1 GPa. The crystal structure of MnP was confirmed by single-crystal X-ray diffraction. Electron probe microanalyses (EDX) indicate a homogeneous distribution of the Mn and P atoms across the whole matrix, suggesting a congruent melting of the compound under high-pressure conditions. Temperature-dependent magnetization measurements reveal the occurrence of two sharp transitions: from paramagnetic to ferromagnetic phase at $T_c = 290.5$ K and, subsequently, a double helimagnetic order at $T_s =$ 44.5 K. The microscopic nature of magnetism of MnP was explored via zero-field $^{31}$P NMR measurements in the FM and the spin-screw AFM state. The relatively large crystals obtained here open up new possibilities for further investigations on this interesting material.

**Acknowledgements**

We would like to acknowledge P. Macchi for his help in performing the structural characterization and J. Hulliger for discussions and helpful comments. The Swiss National Science Foundation is acknowledged for co-funding the single-crystal X-ray diffractometer (R'equip project 206021_128724) and for the financial support provided to NB and TS (Grant no. 200021_169455).

**Table 1.** Details of single-crystal X-ray diffraction data collection and crystal refinement results for MnP.

| | |
|---|---|
| Identification code | shelx |
| Empirical formula | MnP |
| Formula weight | 85.91 g/mol |
| Temperature | 298(2) K |
| Wavelength | Mo K$_\alpha$ (0.71073 Å) |
| Crystal system | Orthorhombic |
| Space group | *Pnma* |
| Unit cell dimensions | $a$ = 5.2510(4) Å,  $\alpha$ = 90° |
| | $b$ = 3.1670(3) Å,  $\beta$ = 90° |
| | $c$ = 5.9098(4) Å,  $\gamma$ = 90° |
| Cell volume | 98.279(14) Å$^3$ |
| Z | 4 |
| Density (calculated) | 5.806 Mg/m$^3$ |
| Absorption coefficient | 13.960 mm$^{-1}$ |
| F(000) | 160 |
| Crystal size | 0.199 × 0.1682 × 0.1255 mm$^3$ |
| $\theta$ range for data collection | 5.194 to 28.160° |
| Index ranges | -6 ≤ h ≤ 6, -4 ≤ k ≤ 4, -7 ≤ l ≤ 7 |
| Reflections collected | 1334 |
| Independent reflections | 136 [$R_{int}$ = 0.0474] |
| Completeness to $\theta$ = 25.000°, % | 100.0 % |
| Absorption correction | Gaussian |
| Max. and min. transmission | 0.268 and 0.114 |
| Refinement method | Full-matrix least-squares on $F^2$ |
| Data / restraints / parameters | 136 / 0 / 14 |
| Goodness-of-fit on $F^2$ | 1.108 |
| Final *R* indices [$I > 2\sigma(I)$] | $R_1$ = 0.0180, w$R_2$ = 0.0438 |
| *R* indices (all data) | $R_1$ = 0.0186, w$R_2$ = 0.0441 |
| Extinction coefficient | 0.077(10) |
| Largest diff. peak and hole | 0.762 and -0.517e.Å$^{-3}$ |

**Table 2.** Position atomic coordinates ( $\times 10^4$ ), in space group *Pnma* at ambient temperature and pressure, and their equivalent isotropic displacement parameters (Å$^2 \times 10^3$) for MnP. *U*(eq) is defined as one third of the trace of the orthogonalized $U^{ij}$ tensor.

| Atom | Wyckoff | x | y | z | *U*(eq) |
|---|---|---|---|---|---|
| Mn1 | 4*c* | 47(1) | 2500 | 1966(1) | 7(1) |
| P(1) | 4*c* | 1880(1) | 2500 | 5686(1) | 7(1) |

**Table 3.** Bond lengths (Å) and angles (°) for MnP.

___________________________________________________________________________________

| | | | |
|---|---|---|---|
| Mn1-P(1)#1 | 2.2849(8) | P(1)#2-Mn1-Mn1#8 | 95.831(16) |
| Mn1-P(1)#2 | 2.3362(6) | P(1)#3-Mn1-Mn1#8 | 149.88(3) |
| Mn1-P(1)#3 | 2.3362(6) | P(1)#4-Mn1-Mn1#8 | 51.371(19) |
| Mn1-P(1)#4 | 2.3841(6) | P(1)#5-Mn1-Mn1#8 | 97.11(2) |
| Mn1-P(1)#5 | 2.3841(6) | P(1)-Mn1-Mn1#8 | 139.810(17) |
| Mn1-P(1) | 2.3998(8) | Mn1#6-Mn1-Mn1#8 | 102.14(2) |
| Mn1-Mn1#6 | 2.7004(3) | Mn1#1-Mn1-Mn1#8 | 100.145(19) |
| Mn1-Mn1#1 | 2.7004(3) | Mn1#7-Mn1-Mn1#8 | 68.54(2) |
| Mn1-Mn1#7 | 2.8121(7) | P(1)#1-Mn1-Mn1#9 | 90.0 |
| Mn1-Mn1#8 | 2.8121(7) | P(1)#2-Mn1-Mn1#9 | 47.328(13) |
| Mn1-Mn1#9 | 3.1670(3) | P(1)#3-Mn1-Mn1#9 | 132.672(13) |
| Mn1-Mn1#10 | 3.1670(3) | P(1)#4-Mn1-Mn1#9 | 48.379(13) |
| P(1)-Mn1#6 | 2.2849(8) | P(1)#5-Mn1-Mn1#9 | 131.621(13) |
| P(1)-Mn1#2 | 2.3362(6) | P(1)-Mn1-Mn1#9 | 90.0 |
| P(1)-Mn1#3 | 2.3362(6) | Mn1#6-Mn1-Mn1#9 | 90.0 |
| P(1)-Mn1#11 | 2.3841(6) | Mn1#1-Mn1-Mn1#9 | 90.0 |
| P(1)-Mn1#12 | 2.3841(6) | Mn1#7-Mn1-Mn1#9 | 124.271(10) |
| P(1)-P(1)#3 | 2.6573(12) | Mn1#8-Mn1-Mn1#9 | 55.729(10) |
| P(1)-P(1)#2 | 2.6573(12) | P(1)#1-Mn1-Mn1#10 | 90.0 |
| | | P(1)#2-Mn1-Mn1#10 | 132.672(13) |
| P(1)#1-Mn1-P(1)#2 | 95.289(13) | P(1)#3-Mn1-Mn1#10 | 47.328(13) |
| P(1)#1-Mn1-P(1)#3 | 95.289(13) | P(1)#4-Mn1-Mn1#10 | 131.621(13) |
| P(1)#2-Mn1-P(1)#3 | 85.34(3) | P(1)#5-Mn1-Mn1#10 | 48.379(13) |
| P(1)#1-Mn1-P(1)#4 | 105.97(2) | P(1)-Mn1-Mn1#10 | 90.0 |
| P(1)#2-Mn1-P(1)#4 | 91.804(16) | Mn1#6-Mn1-Mn1#10 | 90.0 |
| P(1)#3-Mn1-P(1)#4 | 158.73(3) | Mn1#1-Mn1-Mn1#10 | 90.0 |
| P(1)#1-Mn1-P(1)#5 | 105.97(2) | Mn1#7-Mn1-Mn1#10 | 55.729(10) |
| P(1)#2-Mn1-P(1)#5 | 158.73(3) | Mn1#8-Mn1-Mn1#10 | 124.271(10) |
| P(1)#3-Mn1-P(1)#5 | 91.804(16) | Mn1#9-Mn1-Mn1#10 | 180.00(3) |
| P(1)#4-Mn1-P(1)#5 | 83.24(3) | Mn1#6-P(1)-Mn1#2 | 136.274(14) |
| P(1)#1-Mn1-P(1) | 156.93(4) | Mn1#6-P(1)-Mn1#3 | 136.274(14) |
| P(1)#2-Mn1-P(1) | 68.25(3) | Mn1#2-P(1)-Mn1#3 | 85.34(3) |
| P(1)#3-Mn1-P(1) | 68.25(3) | Mn1#6-P(1)-Mn1#11 | 74.03(2) |
| P(1)#4-Mn1-P(1) | 91.103(15) | Mn1#2-P(1)-Mn1#11 | 69.783(12) |
| P(1)#5-Mn1-P(1) | 91.103(16) | Mn1#3-P(1)-Mn1#11 | 123.70(3) |
| P(1)#1-Mn1-Mn1#6 | 150.23(3) | Mn1#6-P(1)-Mn1#12 | 74.03(2) |
| P(1)#2-Mn1-Mn1#6 | 106.39(3) | Mn1#2-P(1)-Mn1#12 | 123.70(3) |
| P(1)#3-Mn1-Mn1#6 | 106.39(3) | Mn1#3-P(1)-Mn1#12 | 69.783(12) |
| P(1)#4-Mn1-Mn1#6 | 54.276(17) | Mn1#11-P(1)-Mn1#12 | 83.24(3) |
| P(1)#5-Mn1-Mn1#6 | 54.276(16) | Mn1#6-P(1)-Mn1 | 70.35(2) |
| P(1)-Mn1-Mn1#6 | 52.83(2) | Mn1#2-P(1)-Mn1 | 111.75(3) |
| P(1)#1-Mn1-Mn1#1 | 56.82(2) | Mn1#3-P(1)-Mn1 | 111.75(3) |
| P(1)#2-Mn1-Mn1#1 | 55.941(19) | Mn1#11-P(1)-Mn1 | 124.20(2) |
| P(1)#3-Mn1-Mn1#1 | 55.941(19) | Mn1#12-P(1)-Mn1 | 124.20(2) |
| P(1)#4-Mn1-Mn1#1 | 137.080(14) | Mn1#6-P(1)-P(1)#3 | 109.37(4) |
| P(1)#5-Mn1-Mn1#1 | 137.080(14) | Mn1#2-P(1)-P(1)#3 | 105.27(4) |
| P(1)-Mn1-Mn1#1 | 100.12(3) | Mn1#3-P(1)-P(1)#3 | 57.01(2) |
| Mn1#6-Mn1-Mn1#1 | 152.95(3) | Mn1#11-P(1)-P(1)#3 | 174.51(3) |
| P(1)#1-Mn1-Mn1#1 | 54.597(19) | Mn1#12-P(1)-P(1)#3 | 101.760(11) |
| P(1)#2-Mn1-Mn1#7 | 149.88(3) | Mn1-P(1)-P(1)#3 | 54.74(3) |
| P(1)#3-Mn1-Mn1#7 | 95.831(16) | Mn1#6-P(1)-P(1)#2 | 109.37(4) |
| P(1)#4-Mn1-Mn1#7 | 97.11(2) | Mn1#2-P(1)-P(1)#2 | 57.01(2) |
| P(1)#5-Mn1-Mn1#7 | 51.371(19) | Mn1#3-P(1)-P(1)#2 | 105.27(4) |
| P(1)-Mn1-Mn1#7 | 139.810(17) | Mn1#11-P(1)-P(1)#2 | 101.760(11) |
| Mn1#6-Mn1-Mn1#7 | 102.14(2) | Mn1#12-P(1)-P(1)#2 | 174.51(3) |
| Mn1#1-Mn1-Mn1#7 | 100.145(19) | Mn1-P(1)-P(1)#2 | 54.74(3) |
| Mn1#1-Mn1-Mn1#8 | 54.597(19) | P(1)#3-P(1)-P(1)#2 | 73.15(4) |

Symmetry transformations used to generate equivalent atoms: #1 *x*-1/2,*y*,-*z*+1/2; #2 -*x*,-*y*+1,-*z*+1; #3 -*x*,-*y*,-*z*+1; #4 -*x*+1/2,-*y*+1,*z*-1/2; #5 -*x*+1/2,-*y*,*z*-1/2; #6 *x*+1/2,*y*,-*z*+1/2; #7-*x*,-*y*,-*z*; #8 -*x*,-*y*+1,-*z*; #9 *x*,*y*+1,*z*; #10 *x*,*y*-1,*z*; #11 -*x*+1/2,-*y*+1,*z*+1/2; #12 -*x*+1/2,-*y*,*z*+1/2

**Table 4.** Anisotropic displacement parameters ($Å^2 \times 10^3$) for MnP. The anisotropic displacement factor exponent takes the form: $-2\pi^2[h^2 a^{*2} U^{11} + ... + 2hk\,a^*b^*U^{12}]$

| Atom | $U^{11}$ | $U^{22}$ | $U^{33}$ | $U^{23}$ | $U^{13}$ | $U^{12}$ |
|---|---|---|---|---|---|---|
| Mn1 | 7(1) | 8(1) | 6(1) | 0 | 0(1) | 0 |
| P(1) | 9(1) | 6(1) | 6(1) | 0 | 0(1) | 0 |

**Figures and captions**

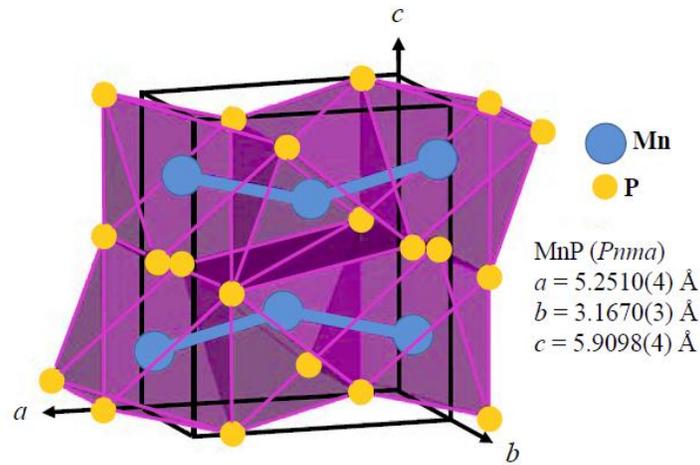

**Figure 1.** (Colour online) 3D view of the MnP-type (*Pnma*, #62, $Z = 4$) orthorhombic structure of the MnP crystal. Mn and P are shown as blue and yellow spheres, respectively. The Mn atoms form zigzag chains along the *a* and *b* axis, which allows the Mn sublattice to be separated into two sublatices (adopted from Ref. [8]).

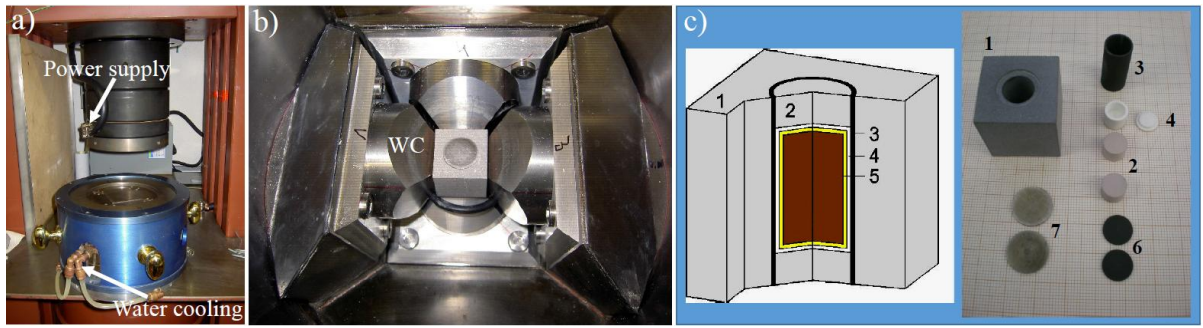

**Figure 2.** (Colour online) (a) Cubic anvil cell apparatus; (b) Top view of the experimental cell: a pyrophyllite cube is placed in the cell core between tungsten carbide (WC) anvils; (c) Cut-away schematic illustration of the sample assembly used for the crystal growth: (1) pyrophyllite cube, (2) pyrophyllite pellets, (3) graphite sleeve, (4) BN crucible, (5) sample in BN crucible, (6) graphite disks, (7) stainless disks.

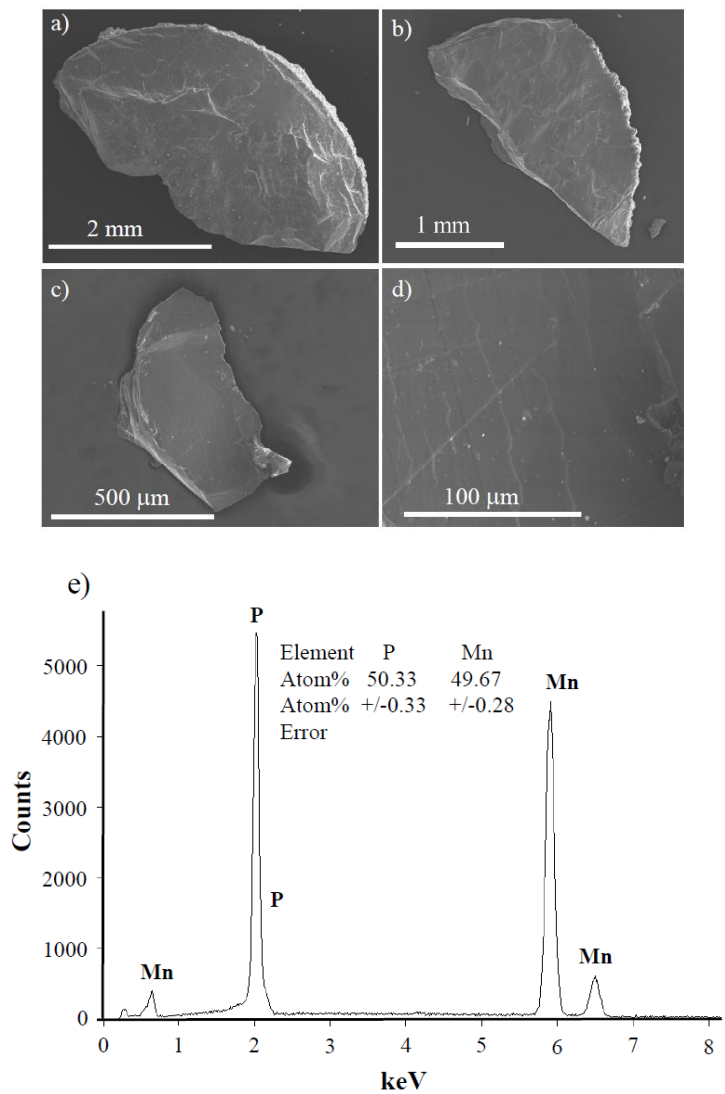

**Figure 3.** SEM micrographs (a-c) of plate-like MnP crystals grown under high pressure. d) A magnified photograph of Fig. 3 (c) shows surface morphology. e) Stoichiometric analysis of a single crystal MnP by Energy-dispersive X-ray spectroscopy (EDX).

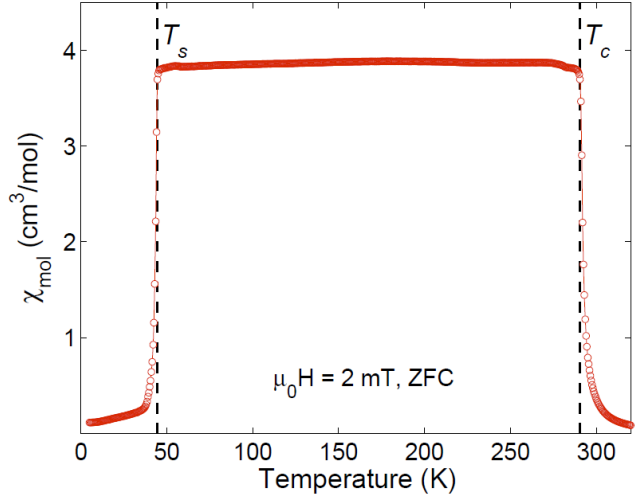

**Figure 4.** (Colour online) Temperature dependence of the molar magnetic susceptibility measured along the *c*-axis at zero pressure. Two evident step-like responses are observed at $T_s$ = 44.5 K, the transition from the FM to the spin-screw AFM state, and at $T_c$ = 290.5 K, the FM transition, both indicated by dashed lines.

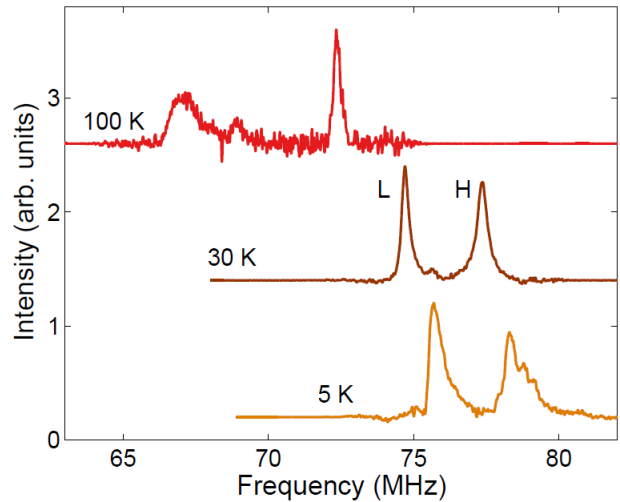

**Figure 5.** (Colour online) Representative $^{31}$P-NMR lineshapes in MnP, measured in the helical (5 and 30 K) and in the FM phase (100 K). In both phases, we observe two peaks, labeled with L (low-frequency) and H (high-frequency), respectively. The line splitting reflects the different magnetic environments for nuclei in the domains (H) and in the domain walls (L) (see text for details).

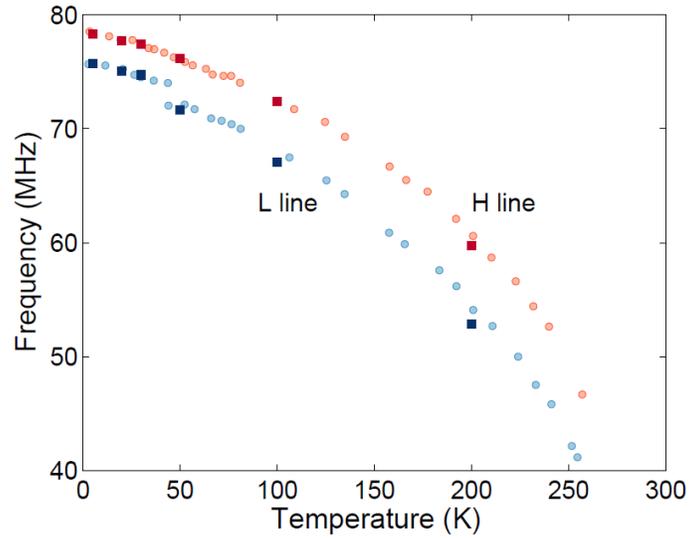

**Figure 6.** (Colour online) Position of zero-field $^{31}$P NMR high- and low-frequency peaks in MnP vs. temperature in the two magnetically-ordered phases. The measured frequencies are consistent with an internal magnetic field $H_{hf}$ = -4.4 T. The current dataset (squares) is in good agreement with previous results (circles) [41].

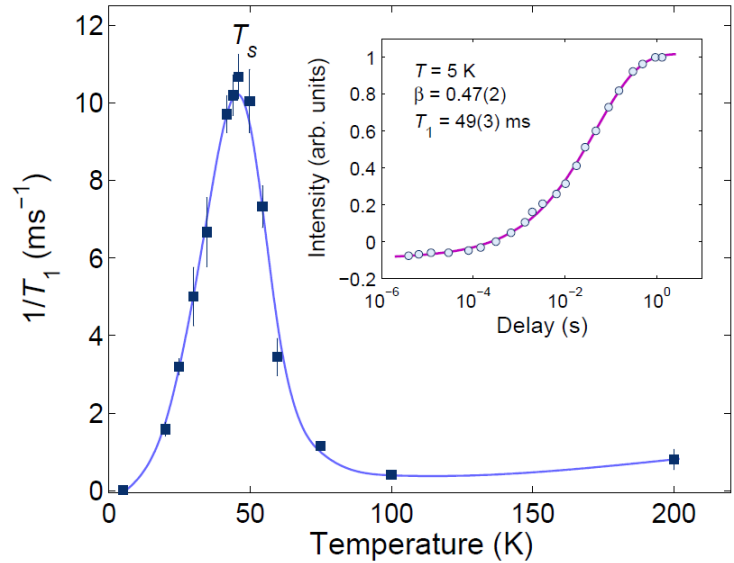

**Figure 7.** (Colour online) $^{31}$P NMR relaxation rate $1/T_1$ as a function of temperature. The prominent peak at $T_s$ indicates the PM-to-helical phase transition. Inset: typical magnetization recovery data (at 5 K) from which $T_1$ is determined by using an exponential fit (continuous line).